# Intersubband Electronic Properties of InAs/GaAs Quantum Dot Molecules with Horizontal Spacer

A. Khaledi-Nasab[1], M. Shahzadeh[2], H. Amouzegar[3], M. Sabaeian[*4]

Physics Department, Faculty of Science, Shahid Chamran University of Ahvaz, Ahvaz, Iran
[1]ali.khaledi1989@gmail.com; [2]shahzadeh.r@gmail.com; [3]hamedamouzegar@yahoo.com; [4*]sabaeian@scu.ac.ir

*Abstract*-In this work the intersubband electronic properties of two laterally coupled dome-shaped InAs/GaAs quantum dots were investigated. The envelope functions and eigenenergies were calculated as function of distance between the dots. The coupling between the dots was studied using transition lifetime between the dots. The results showed that in close distances (smaller than 3 nm) the quantum dots are coupled and by increasing the distance transition lifetime fall down drastically and the dots become uncoupled.

## I. INTRODUCTION

3D confinement of charge carriers in the semiconductor structures have been extensively investigated due to their unique properties and vast variety of applications in different systems such as photonics and opto-electronic devices [1, 2], life sciences [3] and lasers [4]. Such structures can be fabricated using Stranski-Krastanov method in MBE and MOCVD [5]. In this technique, a few number of atomic layers of semiconductors like InAs are deposited on a substrate such as GaAs [5, 6]. Due to lattice mismatch between the growing layer and the base material, the strain effect drives the quantum dots (QDs) towards 3D islands. The unconverted QD material is called wetting layer (WL) [5-7].

Quantum dot molecule (QDM) is formed by coupling two neighbored QDs resulting in formation of electronics and optical states different from those of single QD [8]. Vertical and horizontal (lateral) coupling of QDs have been the subject of research in the past decade [9-12]. Both vertical and horizontal coupled QDs have been proposed for applications in quantum information [13-15].

Since the random nature of self-assembly growth process leads to a size and location distribution of QDs ensemble [16], the investigation of distance-dependent electronic and optical properties of QDs in a QDM is of great important. The distance between QDs in a QDM –so called herein Spacer- has been proved to play an important role in coupling degree of QDs in a QDM [14, 17]. To this objective, Barticevic *et at* proposed a theoretical study of electronic and optical properties of laterally coupled quantum dots under a magnetic field perpendicular to the plane of dots [17]. Bayer *et at* studied the emission of an interacting electron-hole pair in a single QDM as a function of spacer [14].

In this work the intersubband electronic and optical properties of a single dome-shaped InAs/GaAs QDM with horizontal spacer has been investigated. The results will be compared to those of a single QD already presented by Sabaeian *et al.* [5].

## II. THEORY

Schrodinger equation in K.P approximation for a single band electronic structure is given in form of

$$-\frac{h}{8\pi^2}\left[\nabla.\left(\frac{1}{m_e(r)}\nabla\Psi(\vec{r})\right)\right] + V(\vec{r})\Psi(\vec{r}) = E\Psi(\vec{r}) \quad (1)$$

where $h$ is the planck's constant, $m_e$ is the electron effective mass and $\Psi$ is the electron envelope function. The confinement potentials of $V = 0$ and $V = 0.697 eV$ and effective masses of $0.023 m_e$ and $0.069 m_e$ were considered for InAs and GaAs, respectively [5]. Electronic transition dipole moments are calculated by $M_{if} = \left|\left\langle \Psi_f \right| -ex \left| \Psi_i \right\rangle\right|$ where $\Psi_i$ and $\Psi_f$ represent the initial and the final electron's envelope functions. The transition rate and lifetime were calculated in terms of transition dipole moments $\left|M_{if}\right|$, transition frequency $\omega$ and refractive index, $n$, as [18]:

$$W = \frac{1}{\tau} = \frac{\omega^3 n^3}{3\pi c^3 \hbar \varepsilon_0}\left|M_{if}\right|^2 \quad (2)$$

where $c$ is the speed of light, $\varepsilon_0$ is the vacuum permittivity, and n=3.2 is refractive index [5].

Let us consider a linear x-polarized monochromatic electric field propagate along the z direction as $\tilde{E}(z,t) = E_0 \hat{i} e^{i(kz-\omega t)} + C.C$, where $k = N\omega/c$ is the complex propagation constant and $\omega$ is the angular frequency and $N = n + in_i$ is the complex refractive index. Linear absorption is calculated as:

$$\alpha = \frac{2n_i \omega}{c} = \omega \sqrt{\frac{\mu}{\varepsilon_R}} \, \text{Im}\left[\varepsilon_0 \chi^{(1)}(\omega)\right] \quad (3)$$

where

$$\chi^{(1)} = \frac{\sigma}{\hbar} \frac{|M_{21}|^2}{\omega_{21} - \omega - i\gamma_{21}} \quad (4)$$

is the first order susceptibility. In equation above, $\sigma$ is the carrier density and $\gamma_{21}$ is the damping rate for off-diagonal elements of density matrix. These parameters was set to $\sigma = 3 \times 10^{22} \, m^{-3}$ and $\gamma_{21} = 5 \, ps^{-1}$ [5].

III. RESULTS

The cross section of our QDM is shown in figure 1.

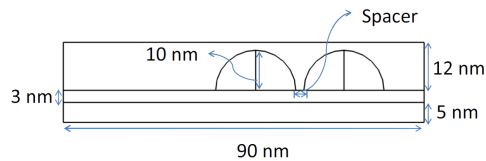

Figure 1. QDM cross section simulated in this work.

The energy eigenvalues corresponding to ground, first, second and third excited states are shown in figures 2.a and 2.b.

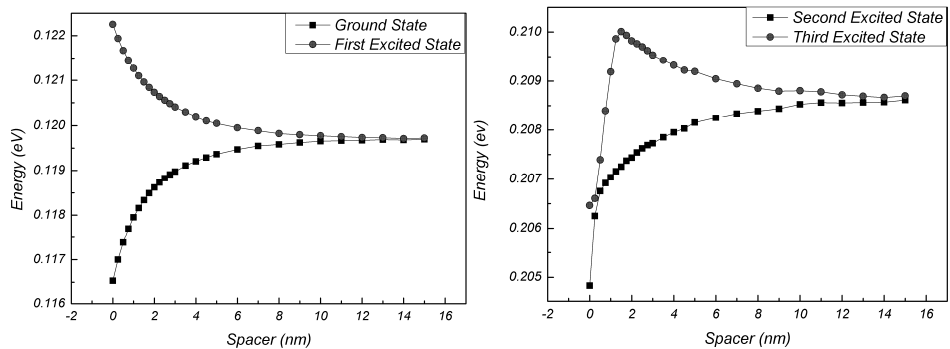

Figure 2. Energy eigenvalues corresponding to ground state and first excited state (a.left) and second excited state and third excited state (b.right).

As the figures show, when the spacer is small, the coupled QDs (QD molecule) behave like a double potential well and the symmetric and antisymmetric envelope functions are similar of ground and first excited states of a single QD (figure 2-a). With increasing the spacer, the mutual interaction between the dots gets weaker and the dots become uncoupled. In this situation the dots act as single QDs and the far QDM energy eigenvalues of ground and first excited states tend to be coincide on ground state energy of a single QD.

Energy eigenvalues of second and third excited states of a QDM are shown in figure 2-b. For a small spacer, the dots envelope function tunnels between the dots through the wetting layer. With increasing the spacer, the coupling effect will become weaker and the tunnelling will be negligible; this is the reason why the energy eigenvalues of second excited state in figure 2-b for small spacer show irregularity (figure 3). With increasing the distance between the dots, the energy eigenvalues of second and third excited states of QDM tend to the first excited state of a single quantum dot.

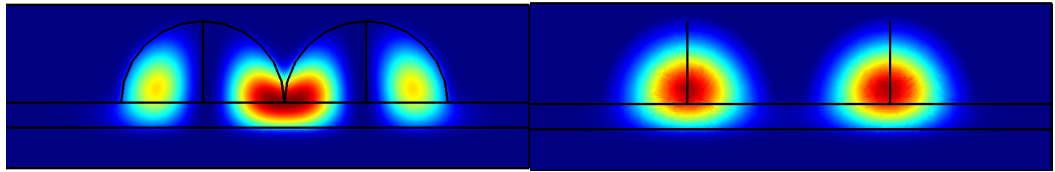

Figure 3. Envelope functions corresponding to second excited state in spacer 0 (left-a) and 5 nm (right-b).

Transition lifetime between the ground and the first excited states with respect to the spacer is shown in figure 4. A transition lifetime of the order of $10^{-4}$ $s$ was calculated for small spacer. Increasing the spacer leads to increasing the transition lifetime and to decoupling. It is worth mentioning that the transition lifetime for a single QD was calculated to be of the order of $10^{-9}$ $s$ [19].

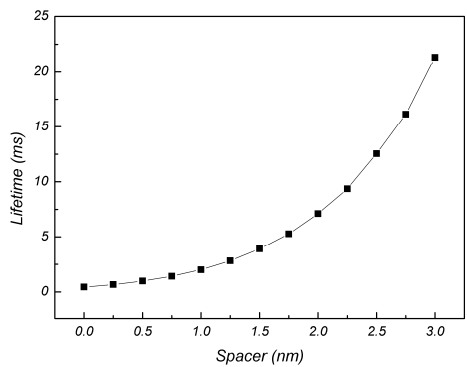

Figure 4. Transition lifetime in QDM with respect to spacer.

Here the reason of relatively long transition lifetime for QDM compared to single QD is discussed. According to figure 2, the energy eigenvalues of a QDM tend to single QD with increasing the spacer and consequently the transition frequency, $\omega$, tends to zero. This effect makes the transition lifetimes to fall down drastically. Since the transition lifetime represents the coupling between the dots, increasing in transition lifetime makes the dots to decouple. Transition lifetimes in more spacers is shown in figure 5.

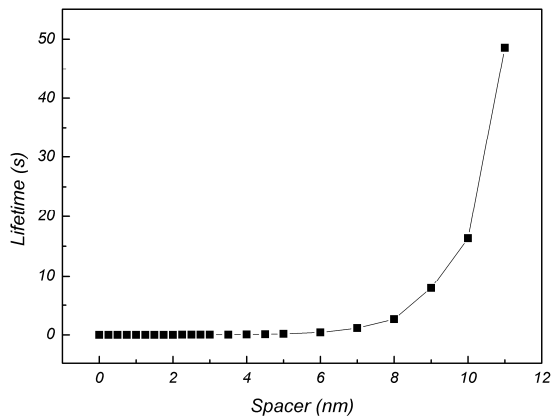

Figure 5. Transition lifetime in QDM with respect to spacer.

For small spacers, the envelope function of neighboured dots lay in the wetting layer and covers two dots, while in sufficiently long distances the envelope function vanishes in the wetting layer connected two dots. The envelope functions in a QDM with zero spacer and 5 nm spacer for ground state are shown in figure 6.

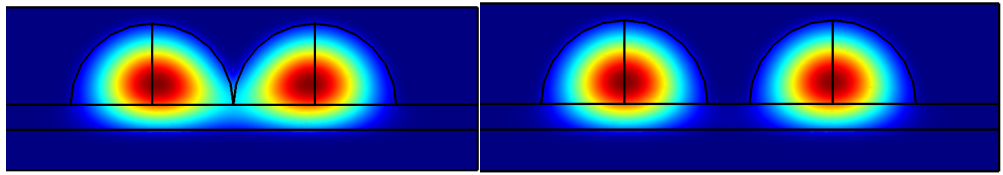

Figure 6. psi^2 corresponding to ground state in a QDM for spacer zero (left) and spacer 5 nm (right).

Linear absorption spectrum is shown in figure 7 corresponding to shoulder to shoulder quantum dots (zero spacer).

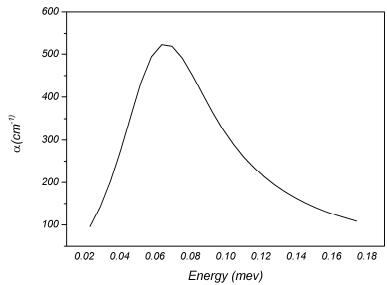

Figure 7. Linear absorption for a QDM corresponding to zero spacer.